\documentclass[
 reprint,superscriptaddress,
 amsmath,amssymb,
 aps,
]{revtex4-2}
\usepackage{amsmath, amsthm, amssymb, amsfonts}
\usepackage{graphicx} 
\usepackage{siunitx}
\usepackage[colorlinks=true, linkcolor=blue, urlcolor=blue, citecolor=blue, anchorcolor = blue]{hyperref} 
\usepackage{color,xcolor}

\usepackage{dcolumn}
\newcolumntype{d}[1]{D{.}{.}{#1}}
\usepackage[utf8]{inputenc}
\usepackage{natbib}
\usepackage{pdfpages}
\usepackage{pgffor} 
\usepackage{xr} 
\usepackage{dcolumn}
\usepackage{bm}
\usepackage{float}
\usepackage[normalem]{ulem}

\makeatletter
\AtBeginDocument{\let\LS@rot\@undefined}
\makeatother
\def\supplementfilename{supplementalmaterial}

\pdfximage{\supplementfilename.pdf}
\def\numbersupplementpages{\the\pdflastximagepages}

\newif\ifarXiv
\arXivtrue 

\usepackage{soul} 
\usepackage[normalem]{ulem}

\begin{document}
\preprint{APS/123-QED}
\title{Coherent All-Optical Radio Frequency Phase Sensing Using Multiphoton Dressing and Interference}
\date{\today}

\author{Hongqiao Zhang}
\affiliation{ Quantum Valley Ideas Laboratories, 485 Wes Graham Way, Waterloo, Ontario N2L 0A7, Canada}

\author{Pinrui Shen}
\affiliation{ Quantum Valley Ideas Laboratories, 485 Wes Graham Way, Waterloo, Ontario N2L 0A7, Canada}

\author{Stephanie M. Bohaichuk}
\affiliation{ Quantum Valley Ideas Laboratories, 485 Wes Graham Way, Waterloo, Ontario N2L 0A7, Canada}

\author{Hanna Lippmann}
\affiliation{ Quantum Valley Ideas Laboratories, 485 Wes Graham Way, Waterloo, Ontario N2L 0A7, Canada}

\author{Harald K\"ubler}
\affiliation{ Quantum Valley Ideas Laboratories, 485 Wes Graham Way, Waterloo, Ontario N2L 0A7, Canada}
\affiliation{5. Physikalisches Institut, Universität Stuttgart, Pfaffenwaldring 57, 70569 Stuttgart, Germany}

\author{James P. Shaffer}
\email{e-mail: jshaffer@qvil.ca}
\affiliation{ Quantum Valley Ideas Laboratories, 485 Wes Graham Way, Waterloo, Ontario N2L 0A7, Canada}
\affiliation{ WaveRyde Instruments, 560 Westmount Road N., Waterloo, ON N2L 0A9, Canada}


\begin{abstract}
Multi-photon dressing and interference in atomic systems is a key to several cutting edge technologies like Rydberg atom radio frequency sensors, clocks and magnetometers because it enables the engineering of atomic properties. Rydberg atom sensors are attracting significant interest because they can be used for applications where it is difficult or impossible to use conventional antennas, opening a number of new opportunities in fields like communications, test and measurement and radar. To date, radio frequency field amplitude detection is well-established in Rydberg electrometry. Phase detection, which is crucial for encoding radio frequency signals, typically requires an external heterodyning field or an atomic closed-loop interferometer. The heterodyne method compromises the intrinsic transparency of the sensor to the radio frequency wave and its inherently broad carrier bandwidth, in addition to increasing its complexity by introducing a local oscillator. In prior theoretical work, aimed at overcoming the disadvantages of the heterodyne method, we theoretically investigated the possibility of using the oscillatory dynamics of an all-optical five-level closed loop to sense the phase and amplitude of the target radio frequency fields. In this work, we experimentally demonstrate the scheme. We determine the coherence time of the loop to be on the order of ms and show that in-phase and quadrature signals can be extracted from a radio frequency signal. 

\end{abstract}

\maketitle
\begin{figure} 
    \centering
    \includegraphics[width=1\columnwidth, 
    trim=10mm 10mm 10mm 10mm, clip]
    {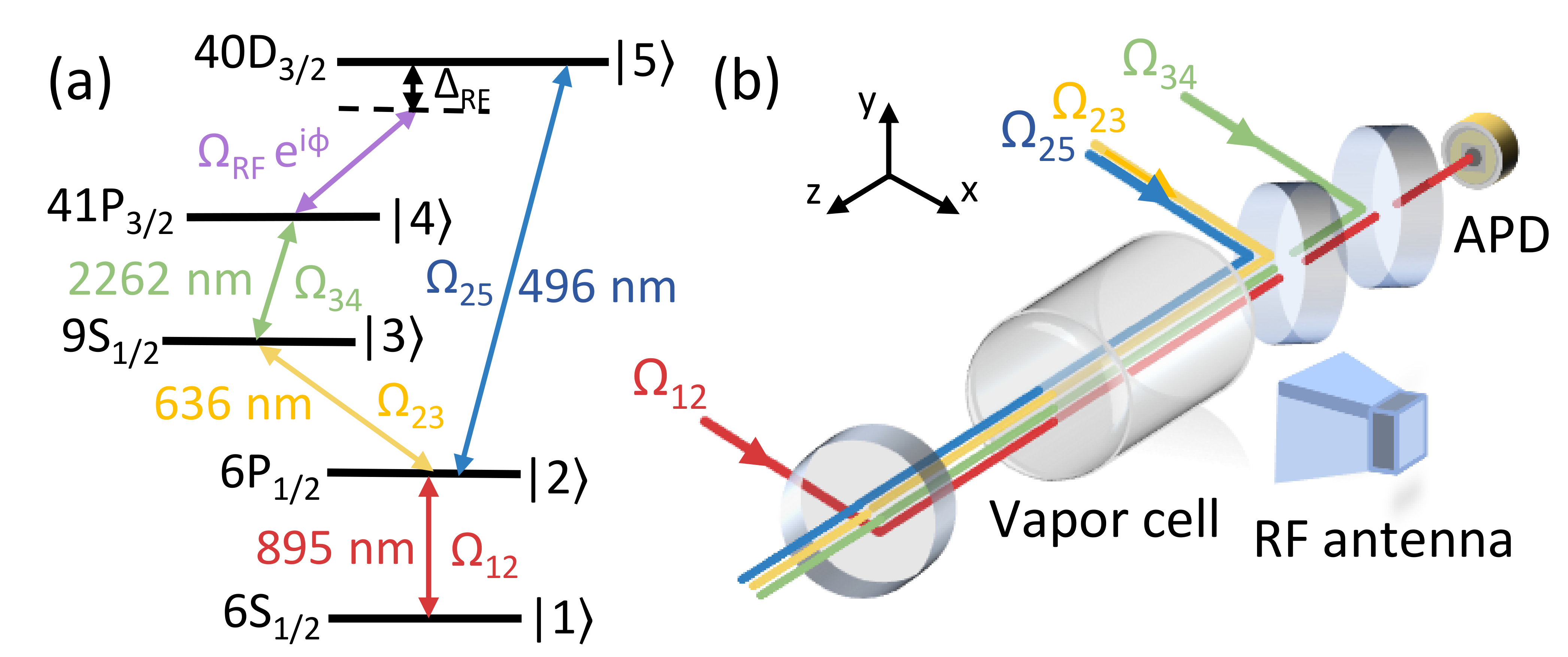}
    \caption{(a) Four photon closed-loop Rydberg rf sensing scheme. An $895\,$nm beam is employed as the probe field. The $636\,$nm, $496\,$nm, and $2262\,$nm beams, together with the target rf field, constitute an optical-rf closed loop. The $10.48\,$GHz rf field is applied with a finite detuning $\Delta_{\mathrm{rf}}$. (b) Experimental setup. The $895\,$nm probe beam counter-propagates co-linearly with three coupling beams through a room-temperature Cs vapor cell ($25\,$mm in length and diameter). The probe transmission is detected by an avalanche photodetector (APD). The rf field is applied via a horn antenna positioned perpendicular to the optical axis. The $1/e^2$ beam radii (and Rabi frequencies, $\Omega$) are $71\,\mu\mathrm{m}$~($2\pi\times5.9\,\mathrm{MHz}$) for $895\,$nm, $886\,\mu\mathrm{m}$~($2\pi\times10\,\mathrm{MHz}$) for $636\,$nm, $270\,\mu\mathrm{m}$~($2\pi\times2.5\,\mathrm{MHz}$) for $2262\,$nm, and $56\,\mu\mathrm{m}$~($2\pi\times3.3\,\mathrm{MHz}$) for $496\,$nm. The Rabi frequency of the rf field is $2\pi\times10.9\,\mathrm{MHz}$.  }
    \label{FIG. 1}
\end{figure}
Multiphoton dressing coherently couples atomic transitions with electromagnetic fields, splitting bare energy levels into multiple sub-levels and inducing quantum interference between transition pathways. This mechanism reshapes atomic properties and drives advancements in quantum technologies, such as Rydberg atom sensors~\cite{fan2015atom}, magnetometers~\cite{budker2007optical}, and atomic clocks~\cite{clockreview}. In radio frequency (rf) sensing, Rydberg atoms have become a new rf sensing technology due to their large transition dipole moments, small dielectric sensor size, and broad carrier bandwidth, making them a highly versatile alternative to conventional antennas across communications~\cite{song2019rydberg,meyer2018digital}, metrology~\cite{sedlacek2012microwave,Holloway,simons2019embedding,robinson2021determining}, and radar~\cite{borowka2024rydberg,10.1063/5.0287757,STEPHANIETRANSIENT}.  
By exciting atoms to high-lying Rydberg states, the strong driving fields establish a split dressed-state manifold. An incident rf field alters the absorption of these dressed states, allowing the rf information to be extracted via electromagnetically induced transparency (EIT) or absorption (EIA)~\cite{sedlacek2012microwave,PhysRevApplied.20.L061004}. 



Thus far, Rydberg atom sensors have predominantly focused on rf electric field strength~\cite{PhysRevApplied.18.014045,PhysRevApplied.18.034030,11123636,kumar2017rydberg,cui2023extending,PhysRevApplied.20.L061004}, because the sensor functions as a square law detector unless an external rf heterodyning field or an atomic looped interferometer is implemented. A heterodyning field acts as a local oscillator and enables phase read out at the expense of the rf equipment necessary to generate it, the carrier bandwidth, the complexity of the system and the electromagnetic transparency of the vapor cell~\cite{simons2019rydberg,8878963,holloway2019detecting,jia2021transfer,liu2022all}. Looped phase read-out has been rarely implemented all-optically, although multi-photon rf transitions have been utilized with the associated disadvantage of non-resonant two-photon absorption~\cite{morigi2002phase,anderson2022optical,berweger2023closed,shylla2018highly,borowka2025optically,PhysRevA.111.053718}. 

To overcome the disadvantages of heterodyne methods, we proposed a four-level closed-loop scheme for all-optical rf phase and amplitude detection in our previous theoretical work~\cite{matthias}. 
By constructing a closed transition loop using optical and rf fields, we induce quantum interference between multiphoton dressed pathways, mapping the rf phase onto a measurable modulation of the probe transmission. 
Here, we experimentally demonstrate this scheme and characterize the coherence time of the loop readout, showing that the in-phase and quadrature components can be extracted from an rf signal.

The excitation scheme is shown in FIG.~1(a). An $895\,$nm probe laser drives the $6S_{1/2}(F=4) \rightarrow 6P_{1/2}(F=3)$ transition. The intermediate state $6P_{1/2}(F=3)$ is coupled to the $40D_{3/2}$ Rydberg state by a $496\,$nm laser. The $636\,$nm laser drives the $6P_{1/2}(F=3) \rightarrow 9S_{1/2}(F=4)$ transition, while a $2262\,$nm laser couples the $9S_{1/2}(F=4) \rightarrow 41P_{3/2}$ transition. A $10.48\,$GHz rf field, with a wavelength of $\lambda_{\mathrm{rf}}= 2.86\,\mathrm{cm}$, drives the $41P_{3/2} \rightarrow 40D_{3/2}$ transition with a finite detuning $\Delta_{\mathrm{rf}}$ and closes the loop. The optical beams are focused into the vapor cell with $1/e^2$ spot radii of $r_{895} = 71\,\mathrm{\mu m}$, $r_{636} = 886\,\mathrm{\mu m}$, $r_{2262} = 270\,\mathrm{\mu m}$, and $r_{496} = 56\,\mathrm{\mu m}$.

The experimental setup is illustrated in Fig.~1(b). 
To cancel the ambient magnetic field and suppress Zeeman shifts of the atomic states, the vapor cell is surrounded by magnetic compensation coils~\cite{PhysRevApplied.20.L061004}. 
All laser beams are linearly polarized and frequency-stabilized to ultralow expansion glass Fabry-Perot cavities using the Pound-Drever-Hall technique. 
The rf field, generated by a signal generator, is coupled into free space via a horn antenna positioned $25\,$cm from the vapor cell.

As discussed in ~\cite{matthias}, 
under the weak probe approximation $\Omega_{12} \ll \Omega_{ij} (ij \neq 1,2)$, the atomic response of the probe laser is given by
\begin{equation}
        \rho_{12}=\frac{C_{12}}{1+\frac{C_{23}C_{32}}{\frac{C_{34}C_{43}}{1+C_{\mathrm{rf}}C'_{\mathrm{rf}}}}+\frac{C_{25}C_{52}}{\frac{C_{\mathrm{rf}}C'_{\mathrm{rf}}}{1+C_{34}C_{43}}}-\frac{2C_{23}C_{34}C_{\mathrm{rf}}C_{52}\mathrm{cos(\phi)}}{1+C_{34}C_{43}+C_{\mathrm{rf}}C'_{\mathrm{rf}}}},
        \label{Eq. (1)}
\end{equation}
where the coherence of a driven transition is $C_{jk}:= i\Omega_{jk}/(\gamma_{k}+2i\Delta_{k})$~\cite{schmidt2024rydberg}. $C_{jk}$($k > j$) and $C_\mathrm{rf}$ denote an upward coherent excitation, while $C_{jk}$($j > k$) and $C_\mathrm{rf}^\prime$  represent a downward stimulated de-excitation. Here, $\Omega_{jk}=\mu_{jk}E_{jk}/\hbar$ is the Rabi frequency of the driving field with $\Omega_{jk}=\Omega_{kj}^*$ and $|\Omega_{jk}|=|\Omega_{kj}|$. $\gamma_{k}$ is the decay rate of state $|k\rangle$. The $\Delta_{k}$ are the multiphoton detunings, given by $\Delta_2 = \Delta_{12}$, $\Delta_3 = \Delta_{12} + \Delta_{23}$, $\Delta_4 = \Delta_{12} + \Delta_{23} + \Delta_{34}$, and $\Delta_5 = \Delta_{12} + \Delta_{25}$. The first two terms in the denominator represent the absorptive responses of the left and right cascaded excitation pathways. They act as two `ladders' constructed within the atom. Each time a new energy level is added to these ladders, the system alternates between destructive interference (EIT) and constructive interference (EIA). This step-by-step dressing process  yields a continued fraction, just as adding a new `rung' to a ladder. The last term is the coherent superposition of a clockwise and a counter-clockwise closed-loop process. These loops can be seen as a four-photon coherence transfer from $|2\rangle$ to $|2\rangle$, which is given by the multiplication of its constituent coherences (e.g., $C_{23}C_{34}C_{\text{rf}}C_{52}$ for the clockwise loop). The right and left handed loop terms are added linearly in the denominator. The two pathways contribute phase terms of $e^{i\phi}$ and $e^{-i\phi}$, respectively, yielding an inverse $\cos(\phi)$ modulation in the probe laser transmission. As can be seen, conventional multiphoton rf sensing depends on the quadratic term $C_\mathrm{rf}C_\mathrm{rf}^\prime$, as described in the first two terms, which is independent of the rf phase. However, the closed transition loop takes advantage of the linearity of the coherence transfer to force the coherent superposition of these transition pathways, inherently preserving the rf phase information.


The loop phase, $\phi$, is
\begin{equation}
    \phi=\Delta_{\mathrm{eff}} \cdot t + k_{\mathrm{eff}} \cdot z + \varphi_{\mathrm{eff}} + k_{\mathrm{eff}} \cdot v \cdot t.\label{Eq. (2)}
\end{equation}
It comprises a detuning-dependent term, a spatial-dependent term, the overall relative phase of the fields, and a Doppler phase shift, with $\Delta_{\mathrm{eff}}=\Delta_{23}+\Delta_{34}+\Delta_{\mathrm{rf}}-\Delta_{25}$, $\varphi_{\mathrm{eff}}=\varphi_{23}+\varphi_{34}+\varphi_{\mathrm{rf}}-\varphi_{25}$, and $k_{\mathrm{eff}}= k_{23}+k_{34}+k_{\mathrm{rf}}-k_{25}$. The signs in these equations depend on geometry \cite{matthias}. Our geometry is shown in Fig.~\ref{FIG. 1}. $v$ is the atomic velocity. In our experimental geometry, the rf field is applied perpendicular to the laser propagation direction, resulting in $k_\mathrm{eff} \approx k_\mathrm{rf}$. In the transverse plane (x-y plane), the spatial phase variation is negligible as the laser beam size is much smaller than $\lambda_{\text{rf}}$. Along the propagation direction, the spatial phase term $k_\mathrm{eff} \cdot z$ is similarly limited, as the interaction region is confined by the tightest Rayleigh range among the laser beams ($z_R = 1.77~\mathrm{cm}$ for the $895~\mathrm{nm}$ probe laser) to a distance comparable to $\lambda_{\mathrm{rf}}$. This geometric confinement ensures that the phase accumulated during spatial averaging does not reach a full period, thereby preventing the signal from being washed out. The perpendicular geometry of the rf field is more straightforward to implement than an end-on geometry and shows that it is possible to design a system where the end-on geometry is not necessary, at least in some cases. The uncompensated $k_\mathrm{eff}$ introduces a velocity-dependent term, causing atoms in different velocity classes to experience different Doppler phase shifts. However, upon integration over the thermal velocity distribution, this term can be observed as a negligible slow amplitude decay rather than a frequency shift. $\varphi_\mathrm{eff}$	is the collective phase of the three lasers and the rf field in the loop. Although fluctuations in the relative phase cannot be entirely eliminated, our experiments demonstrate that the phase information can still be extracted without phase-locking the lasers, which simplifies the practical implementation. According to Eq.~(\ref{Eq. (2)}), when a detuning is applied to one of the coupling fields in the loop, the total loop phase $\phi$ becomes time-dependent. This temporal evolution is imprinted on the probe laser transmission through the $\cos(\phi)$ term, manifesting as detectable oscillations at the loop detuning coming from the modulation of the denominator of Eq.~(\ref{Eq. (1)}). 

\begin{figure} 
    \centering
    \includegraphics[width=1\columnwidth,
    trim=4mm 4mm 4mm 1mm, clip] 
    {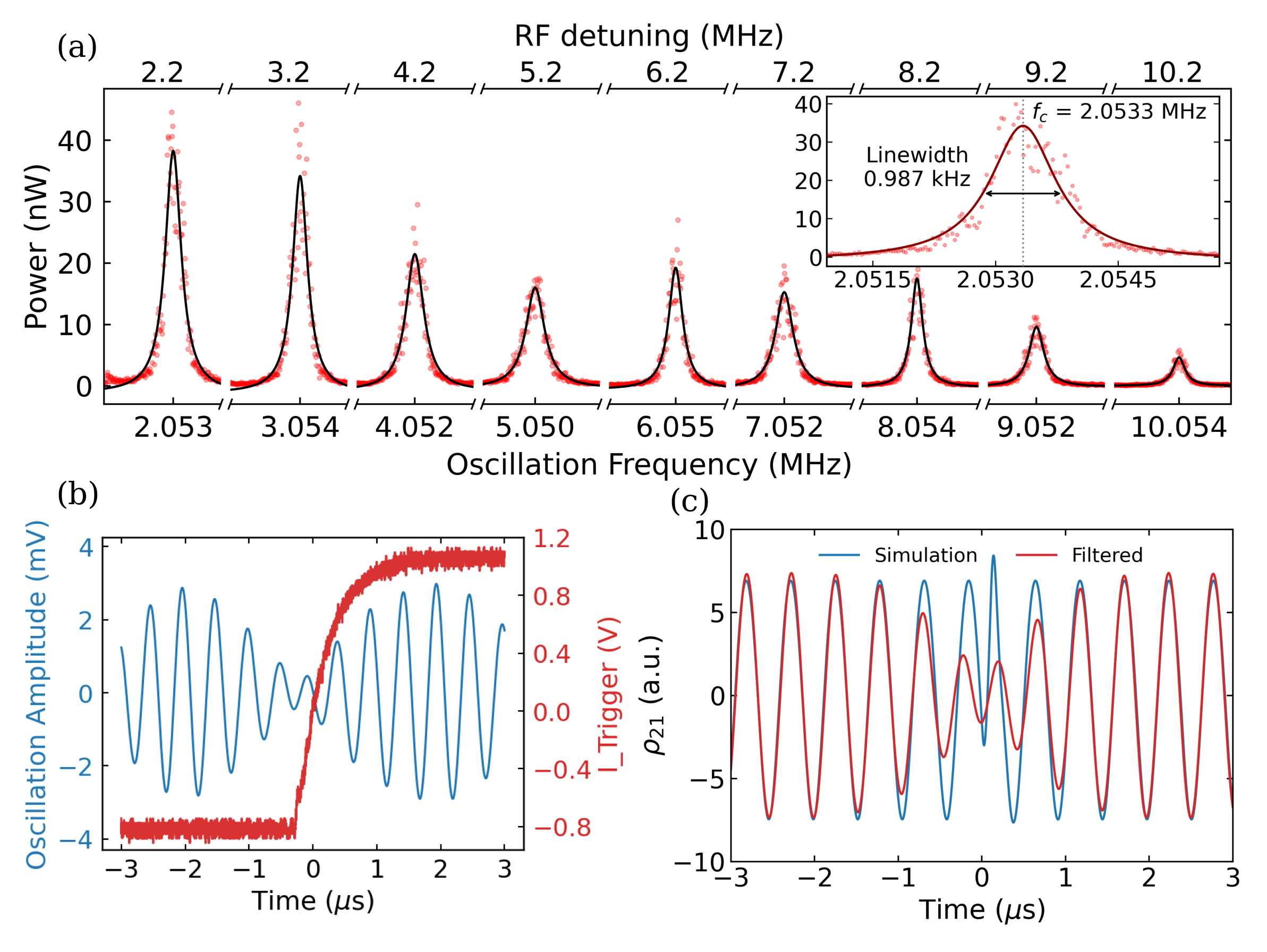}
    \caption{(a) Power spectra of the probe transmission. The main panel shows the transmission peak shifting as the rf detuning is swept from $2.2\,\mathrm{MHz}$ to $10.2\,\mathrm{MHz}$. The inset shows a spectrum centered at $f_{c}=2.0533\,\mathrm{MHz}$ with a Lorentzian full width at half maximum (FWHM) of $0.987\,\mathrm{kHz}$. (b) Oscillatory dynamics and transient response induced by a $5\,$kHz sequence of 180$^\circ$ rf phase jumps. Red curve: In-phase trigger signals from the rf generator. Blue curve: Time evolution of the probe transmission signal with a bandpass filter applied in post-processing ($1.7\mbox{--}2.2\,\mathrm{MHz}$), characterized by an oscillation frequency of $\Delta_{\mathrm{eff}}=1.88\,\mathrm{MHz}$. (c) Simulated transient response. Blue curve: The simulated transmission with a 100$\,$ns rf rise time at $t=0\,\mu$s. The decay rates are $\gamma_2 =2\pi \times 5.98\,\mathrm{MHz}, \gamma_3 =2\pi \times 1.6\,\mathrm{MHz}$, and $\gamma_4 = \gamma_5=2\pi \times1.4\,\mathrm{MHz}$, where the transit time broadening of $2\pi \times 1.4\,\mathrm{MHz}$ is included. All the lasers are on resonance, with a finite rf detuning of $\Delta_{\mathrm{rf}} = 2\pi \times 1.88\,\mathrm{MHz} $. The model includes partial velocity averaging but neglects the spatial averaging. Red curve: The simulated  results using the same 1.7-2.2\,MHz band-pass filter employed in (b).}
    \label{FIG. 2}
\end{figure}

The probe laser transmission is detected by an APD and monitored using a spectrum analyzer. Because the oscillation frequency is defined by the effective detuning of the closed loop, stepping the rf frequency tunes the dynamic response of the system. As shown in FIG.~2(a), the corresponding oscillation frequency precisely tracks the applied rf detuning as it is swept from 2.2 MHz to 10.2 MHz. By applying a Lorentzian fit to the four-photon interference spectra across all measured rf  detunings, we extract a consistent linewidth of $0.94 \pm 0.05~\mathrm{kHz}$ (FWHM). This kilohertz linewidth characterizes the loop coherence time, demonstrating that the multiphoton interference is maintained across a 10\,MHz bandwidth, which is currently limited by the 10\,MHz cutoff of our APD, rather than any breakdown of the atomic response. To reveal the real atomic response, the spectral amplitudes in the main panel have been normalized by the frequency response curve of the APD.

To investigate the oscillation dynamics and the transient response to phase changes, a $5\,\mathrm{kHz}$ sequence of 180$^\circ$ phase jumps is used to modulate the rf carrier. Fig.~2(b) shows the corresponding time evolution of the probe transmission signal after a band-pass filter with $3\,$dB cutoffs at 1.7-2.2\,MHz is applied, revealing a steady oscillation at $1.88\,\mathrm{MHz}$. Due to the finite bandwidth of the filter, the instantaneous phase shift manifests as a distinct amplitude reduction during the rf phase jump. Additionally, a slight amplitude decay is visible near the edges of the observation window, which is primarily attributed to the collective phase noise of the three lasers in the loop. In Fig.~2(c), we numerically solve the Lindblad master equation to model the 180$^\circ$ rf phase step at $t=0\,\mu \mathrm{s}$ with a finite rise time of 100\,ns. When the simulated data is processed through the identical band-pass filter used for the experiment, the resulting signal (red curve) reproduces the transient amplitude reduction feature observed in Fig. 2(b).


\begin{figure}
    \centering
    \includegraphics[width=1\columnwidth]{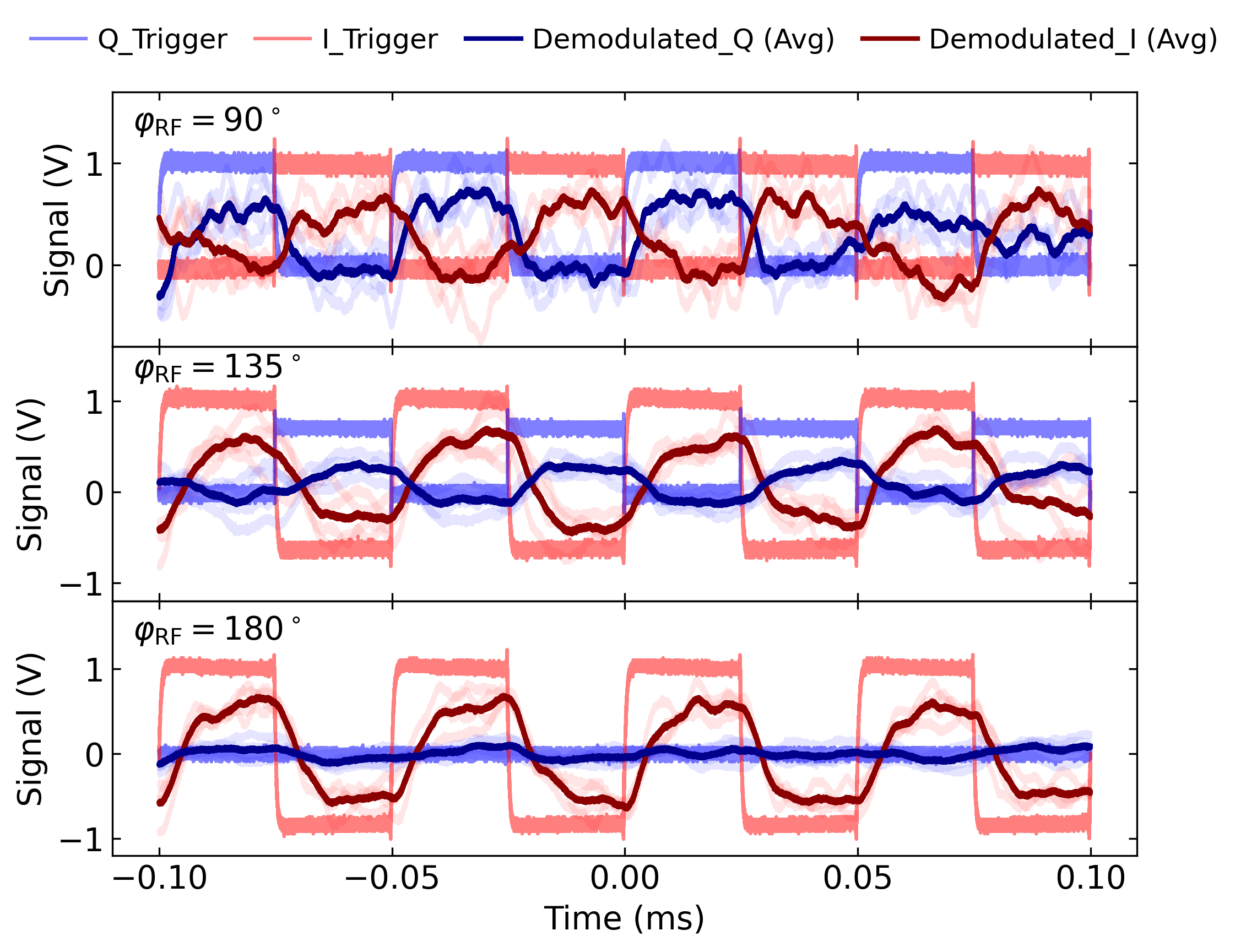} 
    \caption{
    Decoding of phase-modulated signals. The rf field is phase modulated by a $20\,\mathrm{kHz}$ sequence alternating between $0^\circ$ ($(I,Q)=(1,0)$) and target phases of $90^\circ$, $135^\circ$, and $180^\circ$, corresponding to the constellation symbols $(0, 1)$, $(-1/\sqrt{2}, 1/\sqrt{2})$, and $(-1, 0)$, respectively. The probe transmission is processed via a lock-in amplifier for IQ demodulation. Square waveforms: $I$ (red) and $Q$ (blue) trigger signals from the rf signal generator. Solid dark curves: averaged demodulated signals ($N=4$). Solid light curves: single-shot demodulated signals. The observed finite rise and fall times in averaged signals are determined by the bandwidth of the lock-in amplifier's low-pass filter.}
    \label{FIG. 2}
\end{figure}
The oscillatory dynamics carry phase information, enabling the retrieval of discrete phase-modulated signals. To demonstrate the decoding of quadrature encoded signals, we performed three separate experiments applying 20\,kHz phase modulation sequences, Fig.~3. In each sequence, the rf field alternated between a reference phase of $0^\circ$ (corresponding to $(I,Q)=(1,0)$) and a target phase $\phi_{\mathrm{rf}}$ set to $90^\circ$, $135^\circ$, and $180^\circ$, which encode the constellation symbols $(0, 1)$, $(-1/\sqrt{2}, 1/\sqrt{2})$, and $(-1, 0)$, respectively. The probe transmission signal was demodulated using a lock-in amplifier to extract the $I$ and $Q$ components. The accurate retrieval of these discrete phases confirms that the four-photon closed-loop response can be effectively utilized for the recovery of quadrature-encoded information.





We evaluated the detection sensitivity of the loop configuration by measuring the oscillation amplitudes as a function of the applied rf electric field strength. The system exhibits a linear response to the applied field (see Supplemental Materials~\cite{SM}, FIG.~S1).
The threshold field for phase retrival is determined to be $E_{\mathrm{min}}=3.7\,\pm\,2.0\,\mathrm{\mu V/cm}$, corresponding to a sensitivity of $75~\pm~39 \,\mu\mathrm{V/cm/\sqrt{Hz}}$. 
While this is an initial result, it should be noted that since the rf phase is extracted via the probe intensity, any factors that limit the sensitivity in non-looped systems will also affect our system~\cite{PhysRevApplied.20.L061004,schmidt2024rydberg}. Currently, the tightly constrained beam waists required to preserve the spatial phase introduce a substantial transit time broadening ($2\pi \times 1.4\,\mathrm{MHz}$). The sensitivity could be improved by expanding the laser beams, which also increases the number of interacting atoms, provided that spatial phase washout is avoided by using the rf field end-on geometry. Additionally, due to our geometry, the system suffers from a Doppler mismatch, only a fraction of the atoms interact with the lasers. Although configuring the loop fields to counter-propagate against the probe laser could achieve $k_{\mathrm{eff}} = 0$, this is a trade-off, as it would sacrifice the perpendicular geometry for the rf field. The sensitivity is also limited by the laser phase noise within the loop because the random phase fluctuations translate into amplitude noise on the oscillation signal. Phase-locking the lasers would suppress this noise floor and further lower $E_\mathrm{min}$.

In conclusion, we have demonstrated the phase detection of an rf field using a four-photon all-optical closed loop. Through multiphoton dressing, we have engineered an optical interferometer within Rydberg atoms, altering their properties to make the optical response directly dependent on the rf phase. It is important that the phase can be measured using the oscillations of the system, i.e. the phase is determined by a shift of a frequency.  In the experiment, we observed a loop coherence time on the order of milliseconds and retrieved the phase shifts of $90^{\circ}$, $135^{\circ}$, and $180^{\circ}$, demonstrating the capability to decode phase-modulated signals.  By mapping the rf phase onto the optical interference loop, our approach eliminates the need for an external local oscillator required in heterodyning methods. The approach also demonstrates how multi-photon dressing can be used to engineer the properties of atoms so that they can behave in a useful way. 



\acknowledgments

This project was supported by contributions from the Ontario Critical Technologies Initiative (CTI).

%

\clearpage
\ifarXiv
    

\section*{Supplementary material}

# Supplemental Materials to Coherent All-Optical Radio Frequency Phase Sensing Using Multiphoton Dressing and Interference


Hongqiao Zhang,[1] Pinrui Shen,[1] Stephanie M. Bohaichuk,[1]
Hanna Lippmann,[1] Harald Kübler,[1,2] and James P. Shaffer[1,3,*]

[1] Quantum Valley Ideas Laboratories, 485 Wes Graham Way, Waterloo, Ontario N2L 0A7, Canada
[2] 5. Physikalisches Institut, Universität Stuttgart, Pfaffenwaldring 57, 70569 Stuttgart, Germany
[3] WaveRyde Instruments, 560 Westmount Road N., Waterloo, ON N2L 0A9, Canada




## SENSITIVITY ANALYSIS

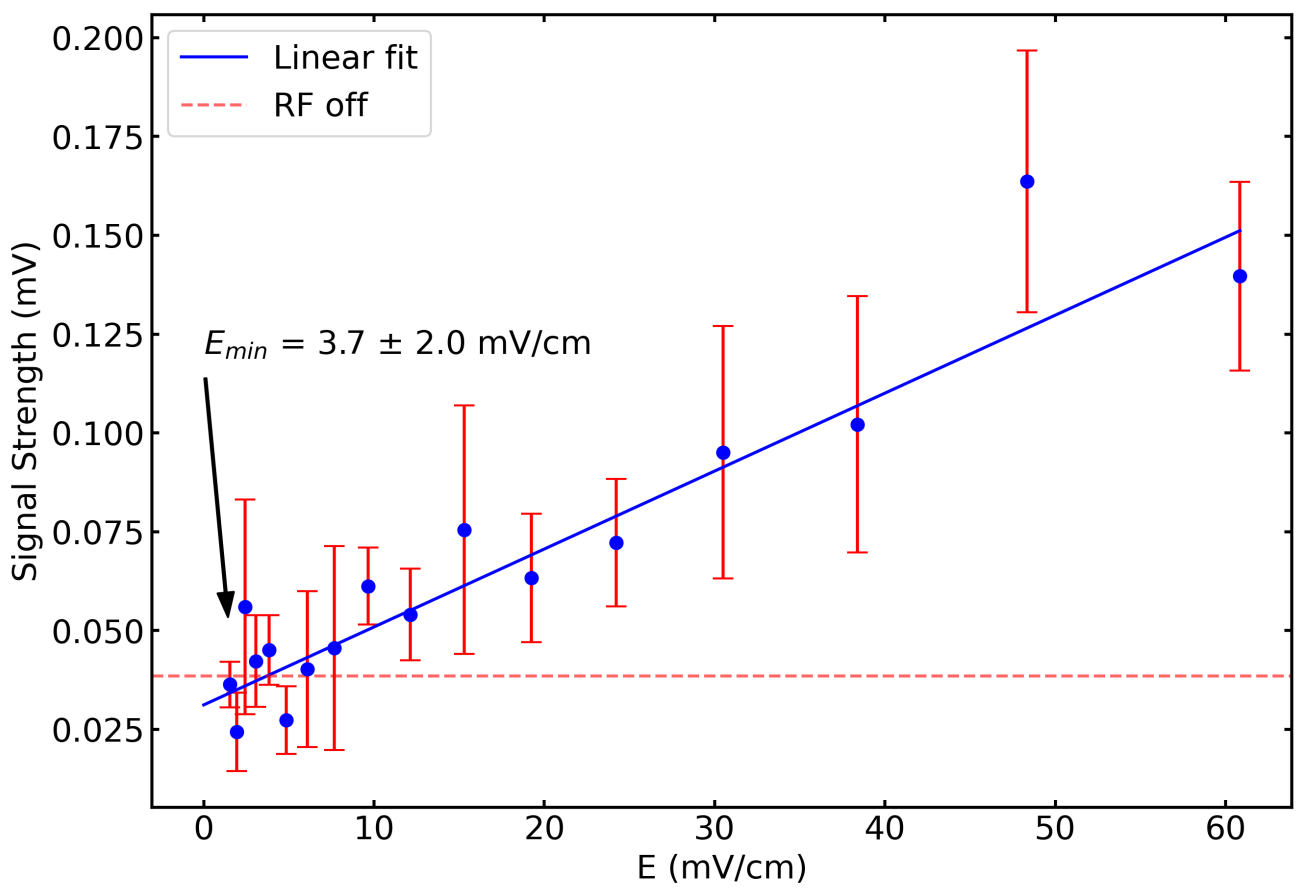


FIG. S1. Sensitivity measurement. The mean oscillation amplitude is plotted as a function of the calibrated rf electric field strength. The data points are obtained over a time window of $\tau = 0.4$ ms, where the error bars represent the standard deviation of the signal amplitude. The solid blue line indicates a linear fit to the data, while the dashed red line marks the system noise floor obtained with the rf field off. The intersection of these lines determines the threshold field for phase retrival of $E_{\mathrm{min}} = 3.7 \pm 2.0$ mV/cm.

We evaluated the detection sensitivity of the loop configuration by measuring the oscillation amplitude as a function of the applied rf electric field strength. Here, the sensitivity is defined as $E_{\mathrm{min}}/\sqrt{B}$, where $E_{\mathrm{min}}$ is the threshold rf electric field required to distinguish oscillation from the noise within a measurement bandwidth $B$. The probe transmission signal was recorded at an oscillation frequency of $\Delta_{\mathrm{eff}} = 1.94$ MHz. To extract the the oscillation amplitude $R(t) = \sqrt{I^2(t) + Q^2(t)}$, the signal was post-processed via a bandpass filter ($1.8 - 2.3\,$MHz) and lock-in demodulation. The extracted amplitude $R(t)$ was time-averaged over a window of $\tau = 0.4$ ms (corresponding to a bandwidth of $B = 2.5$ kHz) across various calibrated electric field strengths. As shown in FIG. S1, the system exhibits a linear response to the applied field. The minimum detetable rf field $E_{\mathrm{min}}$ is determined by the intersection of the linear fit (blue solid line) with the system's noise floor obtained with the rf field off (red dashed line). This yields a threshold field for phase retrival of $E_{\mathrm{min}} = 3.7 \pm 2.0$ mV/cm, corresponding to a sensitivity of $75 \pm 39$ $\mu$V/cm/$\sqrt{\mathrm{Hz}}$.


* e-mail: jshaffer@qvil.ca


\fi

\end{document}